# Touch Survey: Comparison with Paper and Web Questionnaires

Tomoyo Sasao[1], Shin'ichi Konomi[2], Masatoshi Arikawa[2], Hideyuki Fujita[3]
[1]Graduate School of Frontier Sciences, the University of Tokyo, Kashiwa, Japan
[2]Center for Spatial Information Science, the University of Tokyo, Kashiwa, Japan
[3]Graduate School of Information Systems, University of Electro-Communications, Tokyo, Japan
{sasaotomoyo, konomi, arikawa}@csis.u-tokyo.ac.jp, fujita@is.uec.ac.jp

**ABSTRACT**
We developed a prototype of touch-based survey tool for tablets and conducted an experiment to compare interaction patterns of touch-based, PC-based, and paper-based questionnaires. Our findings suggest that a touch-based interface allows users to complete ranking questions easily, quickly, and accurately although it can increase the time to complete a location input task for well-known, prominent locations.

**ACM Classification Keywords**
H.5.2. Information interfaces and presentation (e.g., HCI): User interfaces.

**INTRODUCTION**
Touch devices such as tablets and smartphones provide a ubiquitous platform for collecting various kinds of data from the real world. Researchers and professionals often use traditional paper-based questionnaires as a means to capture relevant information from people. Alternatively, they can use PC/Web-based survey tools that allow for immediate distribution and collection of digital data over the internet. Although such PC/Web-based tools can often be used on smartphones and tablets, a new generation of mobile survey tools is emerging, which actively exploits unique capabilities of mobile devices including location awareness and barcode capture [2][4][6].

To develop a usable and useful input tool on smartphones and tablets, it is essential to understand the characteristics of touch-based interactions with questionnaires. However, there is a relative lack of empirical work uncovering such characteristics. In this context, we developed a prototype of touch-based survey tool and conducted an experiment to compare interaction patterns of tablet-based, PC-based, and paper-based questionnaires. We found that tablet-based questionnaires reduce the time to order items in a ranking question although they can increase the time to complete a location input task for well-known, prominent locations. Also, participants rated the tablet-based interface for a ranking question higher than the PC/Web-based counterpart in terms of ease of use, efficiency, satisfaction, and accuracy of answers, but not ease of recalling answers.

**RELATED WORK**
Many studies have been carried out to examine the effects of administering paper-based and PC/Web-based questionnaires. For example, Kano and Read [3] have shown that children can answer a computer-based questionnaire just as consistently as a paper-based questionnaire. There are similar studies in the fields of medical science and psychology, many of which report no significant difference between the collected data using paper-based and computer-based formats.

Richter et al. [5] examine the usability of a tablet for self-administered questionnaires in a clinical setting. Their study compares tablet questionnaires and paper questionnaires, showing that tablets are a good option for patients with rheumatic diseases to monitor disease activity. Walther et al. [8] compare four electronic data capture (EDC) methods, including the one that exploit tablets, with the conventional paper-based case report forms. Their study shows that EDC methods have the potential to produce similar data accuracy compared to paper-based methods.

Cockburn et al. [1] has investigated performance of basic touch interactions. For example, they show that finger pointing is faster than the stylus and mouse but inaccurate for tapping tasks. Although such basic findings are useful, they are not sufficient to understand interactions with questionnaires. Väätäjä et al. [7] discuss early guidelines for the design of mobile questionnaires without particular attention to the details of touch-based interactions.

**PROTOTYPE DEVELOPMENT**
We have developed a prototype of Touch Survey, a touch-based questionnaire appliance. It is implemented on a 10.1 inch tablet computer (Acer ICONIA TAB W500), and is based on a client-server model. The prototype supports touch-based location input using Google Maps API, audio recording, drawing, and ranking (see Figure 1) in addition to basic single-choice and multiple-choice input. To develop intuitive touch-based interfaces for different types of questions, we not only considered traditional questionnaire formats but also aimed at a design that naturally unfolds questions on a tablet. Each client uploads

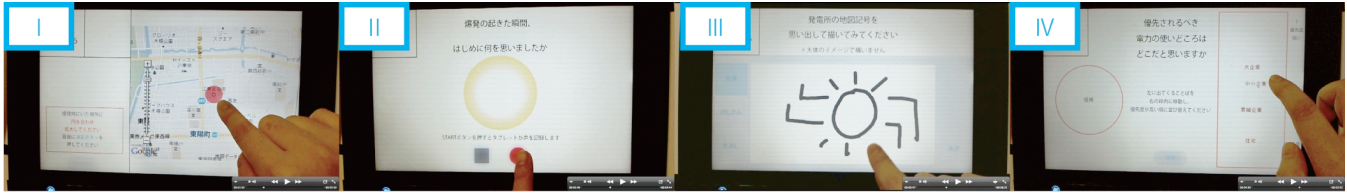

**I. LOCATION INPUT:** (1) Select a prefecture and then a city from a pop-up list to display a map. (2) Perform zoom and drag operations on the map to specify a location.
**II. AUDIO RECORDING:** Touch the recording button and talk freely.
**III. DRAWING:** Draw freely with a finger.
**IV. RANKING:** Items appear in the circle one by one. Drag each item into the rectangle and drop it to sort items incrementally. Ranking order can be modified at any time.

**Figure 1. Tablet-based prototype of Touch Survey**

collected data to a server immediately. The data can be aggregated and displayed in real time. The prototype is not meant as the solution, but a means to understand the characteristics of interactions.

## METHOD

We conducted an experiment to examine if interaction with touch-based, PC/Web-based, and paper-based questionnaires are different. Twenty-two participants (8 women and 14 men), aged 20-50, participated in this study. Participants were recruited from the social networks of the authors. More than 90% of them use a mobile phone or a PC regularly, and 6 are tablet users. Each experiment lasted for approximately 60 minutes, including the time for the participant to answer sample questions and familiarize herself with the survey tools before the experimental phase.

We employed a within-subject design with three trials for each participant, who answered a questionnaire using one of the three different survey tools in each trial: (a) a pen and A4 paper, (b) a online survey tool (SurveyMonkey) on a PC with a 13.3-inch screen, and (c) our prototype on a tablet with a 10.1-inch touch screen. Their order was randomized for participants. Moreover, a different questionnaire content was used in each trial to minimize learning effect. We prepared three questionnaire contents on: (1) experience of an earthquake, (2) saving electricity and (3) sightseeing, and their order was randomized for participants as well. All questionnaire contents have the same structure. That is, each questionnaire content contains five different question types, four of which are illustrated in Figure 1. Questions Q1 and Q2 need to be answered by inputting location (i.e., type I). Q1 asks participants to show a well-known, prominent location (e.g., *"where is the nearest airport?"*) whereas Q2 asks them to show any personally-relevant location (e.g., *"where were you when the earthquake occurred?"*). Q3 needs to be answered by recording audio (i.e., type II), and Q4 by drawing (i.e., type III). Q5 needs to be answered by ranking 15 items (i.e., type IV). Q6 is a multiple-choice question.

To input location information, users simply write down an address on paper, or interact with Google Maps-based interface on a PC and a tablet. On a tablet, users can tap and scroll a map (but without pinching or stretching). A sample ranking question on paper, a PC and a tablet is shown in Figures 2(a), 2(b) and 1(IV), respectively. In Figure 1, ranking items are visualized dynamically in a constrained manner based on a ranking order. In Figure 2(b), the system automatically assigns an unranked item some number from the set of unused numbers, and the display order of the items change automatically based on a ranking order.

We recorded participants' interactions with the survey tools using a video camera, screen video capture software, and system log files. Our prototype recorded all touch events in log files. We analyzed the time spent to perform input operations in detail. After each trial, participants rated perceived ease of use, efficiency, satisfaction, ease of recalling answers, and accuracy of answers for each of the 6 questions, using a likert scale of one to five.

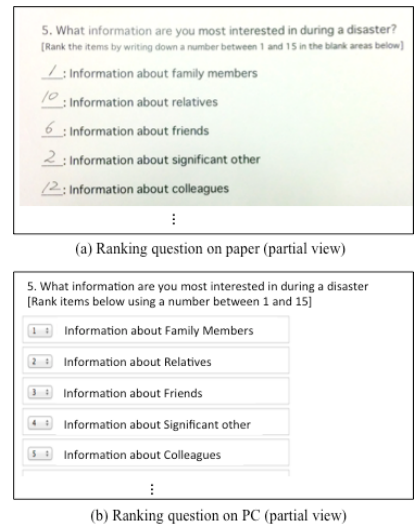

(a) Ranking question on paper (partial view)

(b) Ranking question on PC (partial view)

**Figure 2. Ranking question (Q5) on paper and PC. (The text was translated from Japanese by the authors)**

## FINDING

Table 1 shows the completion time for paper-based, PC/Web-based, and tablet-based questionnaires. A one-way repeated measures ANOVA test using completion time as the dependent variable and survey tools as the independent variable suggests that there is significant difference among the three conditions for questions Q1 [$F(1.465, 30.772)=4.350$, $p<0.05$], Q2 [$F(2, 42)=3.752$, $p<0.05$] and Q5 [$F(2, 42)=3.884$, $p<0.05$]. There are no significant differences for the other questions (Q3, Q4, Q6). Post hoc tests using the Bonferroni correction revealed that the completion time for the ranking questions (Q5) is statistically significantly smaller with the tablet-based survey tool than the PC/Web-based survey tool (p<0.05). In addition, the completion time for the questions for answering prominent locations (Q1) is statistically significantly larger with the tablet than paper (p<0.01).

|     | Paper(Mean,SD) | PC/Web(Mean,SD) | Tablet(Mean,SD) |
| --- | --- | --- | --- |
| Q1 | 37.7 (14.0) | 43.2 (22.3) | 57.8 (32.2) |
| Q2 | 41.8 (20.0) | 41.8 (23.2) | 58.8 (31.4) |
| Q3 | 41.7 (30.7) | 28.2 (17.1) | 29.5 (24.3) |
| Q4 | 34.8 (25.0) | - | 37.9 (23.5) |
| Q5 | 122 (49.8) | 127 (40.4) | 98.1 (32.3) |
| Q6 | 66.2 (17.7) | 64.0 (25.7) | 78.5 (42.6) |

Table 1. Means and standard deviations of completion time in seconds. Q4 was omitted for the PC/Web-based questionnaire that does not allow for drawing input.

|     | Result of Friedman Test |
| --- | --- |
| Ease of use | $\chi^2(2) = 29.662, p = 0.000$ |
| Efficiency | $\chi^2(2) = 22.338, p = 0.000$ |
| Satisfaction | $\chi^2(2) = 32.053, p = 0.000$ |
| Ease of recalling answers | $\chi^2(2) = 14.000, p = 0.001$ |
| Accuracy of answers | $\chi^2(2) = 12.103, p = 0.002$ |

Table 2. Result of Friedman test for subjective ratings of Q5.

|     | Median (IQR) | | |
| --- | --- | --- | --- |
|     | Paper | PC/Web | Tablet |
| Ease of use | 2 (1 to 3) †‡ | 3 (2 to 4) †* | 5 (4 to 5) ‡* |
| Efficiency | 1 (1 to 3) † | 2.5 (2 to 3.25) * | 5 (4 to 5) †* |
| Satisfaction | 2.5 (2 to 3) † | 2.5 (2 to 3) * | 4 (4 to 5) †* |
| Ease recalling answers | 1.5 (1 to 2) † | 2 (1 to 2.25) | 2.5 (2 to 4) † |
| Accuracy of answers | 2 (2 to 3) † | 3 (2 to 3) * | 3 (3 to 4) †* |

Table 3. Median and IQR of subjective ratings of Q5. Pairs of values in each row having the same symbol (†‡*) indicate that the pairs are different statistically significantly (based on Wilcoxon signed-rank test).

### Interaction with the Ranking Questions

Our analysis of the completion time revealed interesting differences for ranking questions (Q5) and for location-input questions (Q1 and Q2). We therefore looked into the detailed interaction patterns with these types of questions.

To conduct an in-depth analysis of the interaction with ranking questions, we divide the ranking task into (1) the first stage, (2) the middle stage and (3) the last stage, based on the number of manipulations. A manipulation is the act of writing down (or inputting) a ranking order of an item in a paper-based (or PC/Web-based) questionnaire, or dragging and dropping an item in a tablet-based questionnaire. Formally, the first, middle, and last stages are defined based on the intervals *[1 .. m/3]*, *[m/3+1 .. 2m/3]* and *[2m/3+1 .. 3m]*, respectively, when a participant has completed a ranking task after *m* manipulations. The number of manipulations varies because users can modify the ranking by performing additional manipulations. Also, the PC/Web-based tool (SurveyMonkey) automatically fills in unspecified ranking orders, and therefore participants can complete a ranking task with a smaller number of manipulations than the number of ranking items.

Figure 3 shows the mean time spent at each stage as well as the mean time spent per manipulation. We can see similar decrease of the mean time as the stage proceeds, for paper-based and PC/Web-based questionnaires. In contrast, there is no such decrease for tablet-based questionnaires: the mean time seems small at all stages. A one-way repeated measures ANOVA test using the spent time as the dependent variable and survey tools as the independent variable suggests that there is significant difference among the three conditions for the first stage [$F(1.403, 29.457)=8.879, p<0.01$] and for the last stage [$F(2, 42)=4.374, p<0.05$]. There are no significant differences for the middle stage. Similarly, there is a significant difference of time spent per manipulation among the three conditions for the first stage [$F(2, 42)=20.092, p<0.01$], middle stage [$F(1.393, 29.243)=19.258, p<0.01$], and the last stage [$F(1.411, 29.636)=12.842, p<0.01$]. Post hoc tests using the Bonferroni correction suggest that tablet-based questionnaires enable statistically significant reduction in the spent time (and time per manipulation) in the first stage.

In addition, there was a statistically significant difference in perceived ease of use, efficiency, satisfaction, ease of recalling answers, and accuracy of answers for Q5, depending on which survey tool was used, as shown in Table 2. Post-hoc analysis with Wilcoxon signed-rank tests was conducted with a Bonferroni correction applied, resulting in a significance level set at $p < 0.017$. The result is shown along with median and IQR values in Table 3.

There was a statistically significant increase in perceived ease of use, efficiency, satisfaction, ease of recalling answers, and accuracy of answers in the tablet-based vs. paper-based questionnaires, and the tablet-based vs. pc-based questionnaires. Moreover, there was a statistically significant increase in perceived ease of recalling answers in the tablet-based vs. paper-based questionnaires.

### Interaction with Location Input Questions

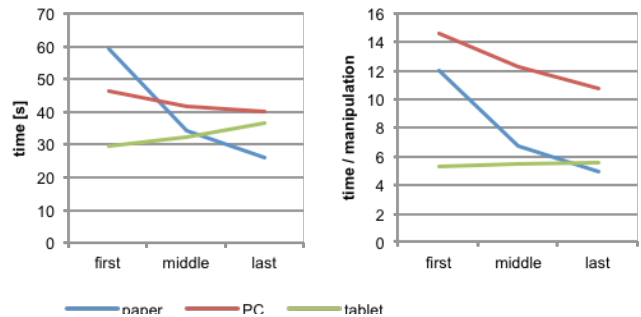

Figure 3. Mean time spent at each stage (left), and mean time spent per manipulation at each stage (right)

We investigated the detailed interaction patterns of tablet-based location input in Q1 and Q2, by counting the number of touch events at each Google Maps zoom level (see Figure 4). The result suggests that people tend to touch the map most at a specific zoom level (i.e., level 14) although patterns are slightly different for different questions. A large number of touch events are often the result of successive map-dragging operations during active search of a location.

**DISCUSSION**
Our study shows that users can complete tablet-based ranking questions faster than PC/Web-based ones. In addition, users can complete paper-based questions about prominent locations faster than tablet-based ones. We now discuss the issues relevant to these results.

**Benefits of Touch-Based Ranking**
The time needed for ranking operations not only include the time for specifying answers but also the time for thinking. It seems that users tend to begin by looking for the 1st items in the ranking when using paper-based or PC/Web-based questionnaires. In contrast, tablet-based questionnaires seem to allow users to think differently, as they support incremental ranking based on relative comparisons. Also, touch-based dragging provides an intuitive means of modifying a ranking order. The smaller completion time, and the increased ease of use, satisfaction and accuracy of answers in tablet-based ranking can be attributed to these aspects of the touch-based interface.

**Different Ways to Answer a Location**
It takes more time to answer a well-known, prominent location using tablet-based questionnaires than paper-based questionnaires, but this result needs to be interpreted carefully because we observed some qualitative differences between these different survey tools. If one can easily remember the name/address of a location, it would be faster to just write it down on paper than navigate to the location by interacting with a digital map.

We observed the following cases in our experiment: (1) some participants explored different locations/answers using zoom and drag operations, (2) a few participants searched the target location starting from their home and navigating along their usual routes using drag gestures. Their answers could have been different had they used paper-based or PC/Web-based questionnaires. Moreover, Figure 4 shows that participants often used a digital map at zoom level 14. This suggests frequent acts of exploration at level 14. The "explorability" with the touch-based interface can be an important benefit.

**CONCLUSION**
This study showed that users can complete tablet-based ranking questions faster than PC/Web-based ones, and the time reduction is particularly significant in the first stage of a ranking task. Tablet-based ranking is rated higher than PC/Web-based ranking in terms of ease of use, efficiency, satisfaction, and accuracy of answers, but not ease of recalling answers. In addition, users can complete paper-based questions about prominent locations faster than tablet-based ones. These results imply the need to provide appropriate survey tools for different types of questionnaires.

There are limitations to our study. First, 14 of the 22 participants are male, which could have potentially biased the result towards men. Second, we only investigated the ranking questions with 15 items; however, the number of items could potentially influence the result. We acknowledge that further work need to be done with a larger number of participants and by considering gender balance, as well as with different numbers of ranking items.

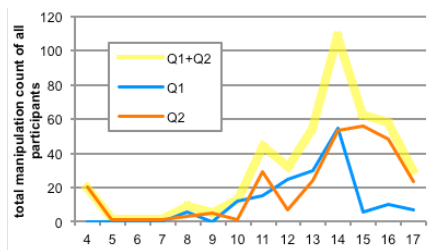

**Figure 4. The number of touch events at different zoom levels. The larger the zoom level, the larger the scale of the map**